\documentclass[11pt]{article}

\def\s2m{$OSp(1/2m,R)$}
\def\s32{$OSp(1/32,R)$}
\def\s64{$OSp(1/64,R)$}
\def\2m{$Sp(2m,R)$}
\def\32{$Sp(32,R)$}
\def\64{$Sp(64,R)$}

\def\s32l{$OSp(1/32,R)_L$}
\def\s32r{$OSp(1/32,R)_R$}
\def\s32lr{$OSp(1/32,R)_{L-R}$}

\newcommand{\eq}{\begin{equation}}
\newcommand{\en}{\end{equation}}
\newcommand{\eqn}{\begin{eqnarray}}
\newcommand{\enn}{\end{eqnarray}}

\begin{document}
\begin{titlepage}
\begin{center}
{\bf AdS/CFT Dualities and the Unitary Representations of Non-compact
Groups and
Supergroups: \\ Wigner versus Dirac} \\
\vspace{1cm}
{\bf Murat G\"{u}naydin\footnote{Work supported in part by the
National Science Foundation under Grant Number PHY-9802510. \newline
e-mail: murat@phys.psu.edu}}   \\
\vspace{.5cm}
CERN \\
Theory Division\\
1211 Geneva 23 , Switzerland \\
\vspace{1cm}
{\bf Abstract}
\end{center}
I review the relationship  between AdS/CFT ( anti-de Sitter / conformal field theory)
 dualities and the general theory of positive energy unitary representations of
non-compact space-time groups and supergroups. I show , in particular, how one can
go from the manifestly unitary compact basis of the lowest weight ( positive energy)
 representations of the conformal group ( Wigner picture) to the manifestly covariant
 coherent state basis ( Dirac picture). The coherent states labelled by the space-time
  coordinates correspond to covariant fields with a definite conformal dimension.
 These results extend to higher dimensional Minkowskian spacetimes as well as generalized
 spacetimes defined by Jordan algebras and Jordan triple systems. The second part of my
 talk discusses the extension of the above results to conformal supergroups of Minkowskian
  superspaces as well as of generalized superspaces defined by Jordan superalgebras. The
  (super)-oscillator construction of generalized (super)-conformal groups
can be given a dynamical realization in terms of generalized
(super)-twistor  fields. \\
\vspace{1cm}

{\it Invited talk presented at the 6th International Wigner Symposium in
Istanbul, Turkey (16-22 August, 1999). To appear in Turkish Journal of
Physics.}

\end{titlepage}
\section{Introduction}
\setcounter{equation}{0}

Since the conjecture of Maldacena \cite{mald}  on the duality between
 the large $\cal N$ limits
of certain conformal field theories (CFT) in $d$ dimensions and the
 superstring theory ,in a certain limit, on the product of $d+1$ dimensional
anti-de Sitter (AdS) spaces with  spheres an enormous amount of research has been
done on AdS/CFT dualities \cite{agmoo}.
 The most studied example of this duality is between the
$ N =4$ super Yang-Mills in $d=4$ and the IIB superstring over $AdS_5 \times S^5$
in the large $\cal N$ limit. In \cite{mgdm} it was pointed out how the conjecture of
Maldacena can be understood  on the basis of some   work
done long time ago on Kaluza-Klein supergravity  theories. Referring to \cite{mgdm}
for details and references to the earlier  work let us recall the salient features
of the earlier work that bear directly on the Maldacena conjecture.
In \cite{gw} the unitary
supermultiplets of the $d=4$ AdS supergroups
$OSp(2N/4,R)$  were constructed and the spectrum of the $S^{7}$ compactification
of  eleven dimensional supergravity was shown to fit into an infinite tower of short
unitary supermultiplets of $OSp(8/4,R)$. The ultra-short singleton supermultiplet of
$OSp(8/4,R)$ sits at the bottom of this infinite tower of Kaluza-Klein modes and
 decouple from the spectrum as local gauge degrees of freedom
\cite{gw}.
However , even though it
decouples from the spectrum as local gauge modes, one can
generate the entire spectrum of 11-dimensional supergravity over $S^7$
by tensoring $p$ copies (``colors'') ($p=2,3,4,\ldots $)  of  singleton
 supermultiplets  and restricting oneself to "CPT self-conjugate "
vacuum supermultiplets. \footnote{ As will be explained below a vacuum supermultiplet
corresponds to a unitary representation whose lowest weight vector is the Fock vacuum.}

The compactification of 11-d supergravity over the four
sphere $S^{4}$ down to seven dimensions was studied in
\cite{gnw,ptn} and its spectrum was shown to  fall into
an infinite tower of unitary supermultiplets
of $OSp(8^{*}/4)$ with the even subgroup $SO(6,2)\times USp(4)$ in \cite{gnw}.
Again the vacuum doubleton supermultiplet of $OSp(8^{*}/4)$ decouples from the
spectrum as local gauge degrees of freedom \footnote{ See the next section for the
distinction between singleton and doubleton supermultiplets.}. It consists of five scalars, four fermions and
a self-dual two form field \cite{gnw}. The entire physical spectrum of 11-dimensional
supergravity over $S^4$ is obtained by simply tensoring an arbitrary number  (colors)
of the doubleton supermultiplets  and restricting oneself to the vacuum
supermultiplets \cite{gnw}.

The spectrum of the $S^{5}$ compactification of ten dimensional
IIB supergravity was calculated in \cite{gm,krv}.
Again the entire spectrum
falls into an infinite tower of massless and massive unitary supermultiplets of
$N=8$ $AdS_{5}$ superalgebra $SU(2,2/4)$ \cite{gm}.
The "CPT self-conjugate" doubleton supermultiplet of
$N=8$ $AdS$ superalgebra  decouples from the physical
spectrum as local gauge degrees of freedom.
By tensoring it with itself
repeatedly and restricting oneself to the $CPT$ self-conjugate vacuum
supermultiplets one generates the entire spectrum of
Kaluza-Klein states of ten dimensional IIB supergravity on $S^{5}$.

The authors of  \cite{gm,mgnm2}  pointed out that the CPT self-conjugate $N=8$  $AdS_5$
doubleton supermultiplet does not have a Poincare limit in five dimensions
and its field theory exists only on the boundary of $AdS_5$ which
can be identified with the $d=4$ Minkowski space .  Furthermore, they pointed
out, for the first time, that  the doubleton field
theory of $SU(2,2/4)$ is the conformally invariant
$N=4$ super Yang-Mills theory in $d=4$.
Similarly, the singleton supermultiplet of $OSp(8/4,R)$ and the doubleton
supermultiplet of $OSp(8^*/4)$ do not have a Poincare limit in $d=4$ and
$d=7$, respectively, and their field theories are conformally invariant theories
in one lower dimension \footnote{ see \cite{mgdm} for references }.
Thus we see that at the level of physical states the proposal of Maldacena
is perfectly consistent with  the above
mentioned results if we assume that the spectrum of the
superconformal field theories fall into ("$CPT$ self-conjugate" )
vacuum supermultiplets. Remarkably, this is equivalent to assuming that
the spectrum consists of ``color'' singlet supermultiplets. !

\section{ Massless and Massive Supermultiplets of Anti-de Sitter Supergroups}
\setcounter{equation}{0}

The Poincar\'{e} limit of the remarkable representations (singletons)
 of the $d=4\  AdS$ group
$SO(3,2)$  discovered by Dirac \cite{pad} are known to be singular \cite{cf1}.
 However, the tensor product of
 two singleton unitary irreducible representations of
$SO(3,2)$ decomposes into an infinite set of massless unitary irreducible
 representations which do have a
smooth Poincar\'{e} limit \cite{cf1,gs,mg81}.
 Similarly, the tensor product of two
singleton supermultiplets of $N$ extended $AdS_4$ supergroup $OSp(N/4,R)$
decomposes
into an infinite set of massless supermultiplets which do have a Poincar\'{e}
limit in five dimensions \cite{mg81,gw,gh,mg89}.
 The $AdS$ groups $SO(d-1,2)$ in higher
 dimensions than four
 that do admit  supersymmetric extensions  have doubleton representations only.
 The doubleton supermultiplets of extended $AdS$ supergroups in $d=5\
 (SU(2,2/N))$ and
$d=7\ (OSp(8^*/2N))$ share the same remarkable features of the singleton
supermultiplets of $d=4$ $AdS$ supergroups i.e  the tensor product of any two
doubletons decompose into an infinite set of massless supermultiplets
\cite{gnw,gm,mg89,gmz1,gmz2,mgst}.
 In $d=3$ the $AdS$ group $SO(2,2)$ is not simple and is isomorphic to
$SO(2,1)\times SO(2,1)$. Since each $SO(2,1)$ factor can be extended to a simple
superalgebra with some internal symmetry group  one has a rich variety of $AdS$
supergroups in $d=3$ \cite{gst}. Since locally we have the isomorphisms
$SO(2,1)\approx
SL(2,R)\approx SU(1,1)\approx Sp(2,R)$ the $AdS$ supergroups in $d=3$ (and
hence in $d=2$) admit singleton representations \cite{gst}.

Since the Poincar\'{e} mass operator is not an invariant ( Casimir) operator of
the $AdS$ group   the following definition of a massless
representation (or supermultiplet) of an $AdS$ group (or supergroup)
was proposed in \cite{mg90}:

{\it A representation (or a supermultiplet) of an $AdS$ group (or supergroup) is
 massless if it occurs in the decomposition of the tensor product of
two singleton or two doubleton representations (or supermultiplets).}

The tensor product of more than two copies of the singleton or doubleton
supermultiplets of $AdS$ supergroups decompose into an infinite set of massive
supermultiplets in the respective dimensions as
has been amply demonstrated within the Kaluza-Klein supergravity theories
\cite{gnw,gw,gm} and more recently \cite{gmz1,gmz2,mgst} .
 A noncompact group that admits only
doubleton representations can always be embedded in a larger noncompact group
that admits singleton representations. In such cases
 the singleton representation of the
larger group decomposes , in general, into an infinite tower of doubleton
 representations of
the subgroup.

\section{Unitary Lowest Weight Representations of Noncompact Groups}
\setcounter{equation}{0}

A representation of a non-compact group is said to be of the lowest weight type if
 the spectrum of at least one of its generators is bounded from below within
 the representation space.
A non-compact simple group G admits unitary lowest weight representations (ULWR)
 if and only if its quotient space G/H
with respect to its maximal compact subgroup H is an hermitian symmetric space
\cite{hc}. Thus the complete list of simple  non-compact groups G that admit ULWRS's
follows from the list of irreducible hermitian symmetric spaces \cite{hc} which we
give below:

\begin{table}[ht]
  \caption{~}
\begin{center}

\begin{displaymath}
\begin{array}{|c|c|}
\hline
G & H \\
\hline
& \\
SU(p,q) & S(U(p)
\times U(q)) \\ \hline
& \\
Sp(2n,R)& U(n) \\ \hline
& \\
SO^*(2n)& U(n) \\ \hline
& \\
SO(n,2)& SO(n)
\times SO(2) \\ \hline
& \\
E_{6(-14)}& SO(10)\times
U(1) \\ \hline
& \\
E_{7(-25)}& E_6
\times U(1)     \\
& \\
\hline
\end{array}
\end{displaymath}
\end{center}
\end{table}

The Lie algebra $g$ of a non-compact group G that admits ULWR's has a 3-grading
 with respect to the Lie algebra $h$ of its maximal
compact subgroup H i.e

\begin{eqnarray}
g=g^{-1}\oplus g^0\oplus g^{+1}
\end{eqnarray}
where $g^0=h$ and we have the formal commutation relations
\begin{eqnarray}
{[}g^{(m)},g^{(n)}{]}\subseteq g^{(m+n)}\hspace{2.0cm}m,n=\mp 1,0\nonumber
\end{eqnarray}
and $g^{(m)}\equiv 0$ for $\vert m\vert >1.$
\vskip 0.3 cm
In \cite{gs} the general  oscillator construction of unitary lowest weight representations (ULWR)
 of non-compact groups  was given. Particular cases of the oscillator
 construction for certain representations of some special groups such as
 $SU(1,1)$ and $SU(2)$ had previously appeared in the physics literature.

To construct the ULWR's one first realizes the generators of the noncompact group $G$
as bilinears of bosonic oscillators transforming in a certain representation of $H$.
Then in the corresponding Fock space ${\cal F}$
 one chooses  a set of states $|\Omega >$, referred to as the
"lowest weight vector" (lwv), which transforms irreducibly under  $H$ and which
 are annihilated by the generators belonging to the
$g^{-1}$ space. Then by acting on $|\Omega >$ repeatedly with the
generators belonging to the $g^{+1}$ space one obtains an infinite set of states
\eq
|\Omega >,\ \ g^{+1} |\Omega >,\ \ g^{+1}g^{+1} |\Omega >,...
\en
This set of states  forms the basis of an irreducible unitary lowest weight
 representation of $g$. (The
 irreducibility of the representation of $g$ follows from the
irreducibility of  the lwv $ |\Omega >$  under $h$.  The
 bosonic oscillators $a_i(r)$
satisfy the canonical commutation relations
\begin{equation}
\begin{array}{c}
~\\
{[}a_{i}(r),a^{j}(s){]} = \delta_{i}^{j} \delta_{rs}\hspace{2.0cm}i,j=1,...,n
\\
{[}a_{i}(r),a_{j}(s){]} = 0  \hspace{2.0cm}     r,s=1,...,p
~
\end{array}
\end{equation}

where the upper indices $i,j,k,... $ are the indices in the representation $R$ of $h$ under
 which the oscillators transform and $r,s.. =1,2,..p$ label the different sets of oscillators.
 We denote the creation (annihilation) operators with
upper (lower) indices $i,j,..$, respectively :

\begin{eqnarray}
a_i(r)^\dagger\equiv a^i(r) \nonumber
\end{eqnarray}
\\
Generally, $R$ is the fundamental representation of  $h$ and we
 shall refer to $p$ as the  number  of colors. The generators are  color
singlet bilinears but the lowest weight vector $|\Omega >$ and hence the infinite tower
of vectors belonging to the corresponding ULWR can carry color. Depending on the
non-compact group the minimal number $p$ of colors required to realize the generators
can be one or two. If  $p_{min}=1$, we shall call the corresponding unitary irreducible
 representations singletons and if
$p_{min}=2$, they will be referred to as doubletons \cite{gnw}.
 The   non-compact groups $Sp(2n,R)$ admit singleton unitary irreducible representations
 \cite{gs,gw,gh,mg89} while the groups
$SO^*(2n)$ and $SU(n,m)$ admit doubleton unitary irreducible representations
 \cite{gs,gnw,gm,mgrs,mg89}. We should note that the ``remarkable representations'' of
 the  four dimensional $AdS$ group $SO(3,2)$ with the  covering group $Sp(4,R)$
 discovered by Dirac
\cite{pad} are simply the singletons.
 While when $p_{min}=1$ for a given non-compact group there exist only {\it two}  singletons ,
 one finds
infinitely many doubletons for $p_{min}=2$.
 The two singletons  of  $Sp(4,R)$  can be associated with  spin zero and spin
$\frac{1}{2}$ fields. On the other hand the $d=7\ AdS$ group
$SO^*(8)=SO(6,2)$ admits infinitely many doubletons corresponding to
fields of arbitrarily large spin \cite{gnw,mgst}. However, we should
 note that the doubleton fields are not of
the form of the most general higher spin fields in $d=7$. Their decomposition
with respect to the little group $SU(4)\equiv Spin(6)$ in $d=7$ correspond to
those representations of $SU(4)$ whose Young-Tableaux have only one row \cite{gnw,mgst}. Whereas
the general massive higher spin fields correspond to the representations of the
little group with arbitrary Young-Tableaux.

If one replaces the bosonic oscillators  with fermionic
ones, then the above construction leads to  the unitary representations of the compact
 forms of the
corresponding groups.  One finds that the compact $USp(2n)$ admits
doubleton (unitary irreducible) representations (finitely many) while the group $SO(2n)$ admits
 two singleton  (unitary irreducible) representations \cite{mg90}.
 The singletons of $SO(2n)$ are the two spinor
 representations.
In general the compact group $USp(2n)$ admits $n$ non-trivial doubleton  representations.
 For $USp(4)$ they are the spinor representation (\underline{4}) and the adjoint
 representation (\underline{10}).
The two singletons (spinors)  of $SO(2n)$ combine to form
the unique singleton (spinor representation) of $SO(2n+1)$.

\section{ Unitary Lowest Weight Representations of Noncompact Supergroups}
 \setcounter{equation}{0}
The extension of the oscillator method to the construction of the ULWR's of
non-compact supergroups with a three-graded  structure with respect to a maximal
compact subsupergroup was given in \cite{bg} \footnote{  A non-compact supergroup
is defined as a supergroup whose even subgroup has a non-compact subgroup.}
. This method was further developed
and applied to space-time supergroups and Kaluza-Klein supergravity theories
in the eighties \cite{gnw,gw,gm,gst} . The general construction of the ULWR's of
 the noncompact supergroup $OSp(2n/2m,R)$ with the even subgroup $SO(2n)\times Sp(2m,R)$ was
 studied in \cite{gh} and the ULWR's of $OSp(2n^*/2m)$ with the even subgroup $SO^*(2n) \times
USp(2m)$ in reference \cite{mgrs}. More recently a detailed study of the unitary
 supermultiplets of the supergroups $SU(2,2/4)$ and of $OSp(8^*/4)$ relevant to AdS/CFT
dualities in M-theory was given in \cite{gmz1,gmz2} and \cite{mgst} , respectively.

Consider now the  Lie superalgebra $g$ of a non-compact supergroup $G$ that
has a 3-graded structure with respect to a compact subsuperalgebra $g^0$ of maximal rank
\begin{eqnarray}
g=g^{-1}\oplus g^0\oplus g^{+1}\nonumber
\end{eqnarray}
To
construct the ULWR's of $g$ we first realize its generators as bilinears of a set of
superoscillators $\xi_A(\xi^A)$ whose first $m$ components are bosonic and
the remaining $n$ components are fermionic
\begin{eqnarray}
\xi_A(r)=\left(\begin{array}{cc}
a_i(r)\\
\alpha_\mu(r)\end{array}\right)\ \ \xi^A(r)=\left(\begin{array}{cc}
a^i(r)\\
\alpha^\mu(r)\end{array}\right)\nonumber
\end{eqnarray}
\\
\hspace*{4.5cm}$i,j=1,...,m\ \ ;\ \ \mu,\nu=1,...,n$\\
\\
\hspace*{4.5cm}$r,s=1,...,p.$\\
\\
which satisfy the supercommutation relations
\begin{eqnarray}
{[}\xi_A(r),\ \xi^B(s)\}=\delta_A\,^B\delta_{rs}
\end{eqnarray}
where [\ ,\ \} means an anti-commutator for any two fermionic oscillators and a
commutator otherwise. Furthermore we have

\begin{eqnarray}
{[}\xi_A(r),\ \xi_B(s)\}=0={[}\xi^A(r),\ \xi^B(s)\}
\end{eqnarray}

Generally the operators belonging to the $g^{-1}$ and $g^{+1}$ spaces are
realized as super di-annihilation and di-creation operators respectively. Consider  now
 a  lowest weight vector $|\Omega >$, that transforms irreducibly under
$g^0$ and is annihilated by $g^{-1}$ operators. Acting on $|\Omega>$ with the $g^{+1}$
 operators
repeatedly one generates an infinite set of states that form the basis of a  ULWR of $g$\\

\begin{eqnarray}
g^{-1}|\Omega >=0\ ,\ \ \ \ g^0|\Omega >=|\Omega'>\hspace{2.0cm}
\nonumber\\
\ \\
\{{\rm ULWR}\}\equiv\{|\Omega >,\ g^{+1}|\Omega >,\ g^{+1}g^{+1}
|\Omega >,...\}
\end{eqnarray}
The resulting ULWR is uniquely labelled by $|\Omega>$.
A supergroup $g$ admits singleton or doubleton unitary irreducible representations
depending on whether $p_{min}=1$ or $p_{min}=2$, respectively. For example
the non-compact supergroup $OSp(2n/2m,R)$ with even subgroup $SO(2n)\times
Sp(2m,R)$ admits singleton representations. The non-compact supergroup
$OSp(2n^*/2m)$ with even subgroup $SO(2n)^*\times USp(2m)$ admits doubleton
 representations, as does the supergroup $SU(n,m/p)$ with even subgroup
$S(U(n,m)\times U(p))$. There exist only two irreducible singleton
supermultiplets of the non-compact supergroup $OSp(2n/2m,R)$ \cite{gw,gh}. On the
other hand, the supergroups $OSp(2n^*/2m)$ and $SU(n,m/p)$ admit infinitely
many irreducible doubleton supermultiplets \cite{gnw,mgrs,gm,gmz1,gmz2}.

In contrast  to the situation with noncompact groups, not all noncompact
supergroups that have ULWRs admit a three grading with respect
to a compact subsupergroup of maximal rank. The method of \cite{bg} was
 generalized to the
case when the noncompact supergroup admits a 5-grading with respect to a
compact subsupergroup of maximal rank in \cite{mg88}.
 For example, the superalgebra of  $OSp(2n+1/2m,R)$
admits a 5-grading with respect to its compact subsuperalgebra $U(n/m)$
 , but it does not admit a three grading with respect to a compact subsuperalgebra of
 maximal rank for general $n$ and $m$.
All  finite dimensional non-compact supergroups do admit a 5-grading with respect to a
 compact
subsupergroup of maximal rank  \cite{mg88}.

\begin{eqnarray}
g=g^{-2}\oplus g^{-1}\oplus g^0\oplus g^{+1}\oplus g^{+2}
\end{eqnarray}

\section{Generalized space-times  defined by Jordan  algebras}
\setcounter{equation}{0}

\subsection{ Generalized Rotation, Lorentz and Conformal Groups}

The twistor formalism in  four-dimensional space-time $(d=4)$
leads naturally to the representation of four vectors  in terms
of $2\times 2$  Hermitian matrices over the field of complex
numbers ${\mathbf C}$.  In particular, the coordinate four vectors $x_{\mu}$
can  be represented  as :
\eq
x=x_{\mu}\sigma^{\mu}
\en

Since the Hermitian matrices over the field of complex
numbers close under the symmetric anti-commutator product
we can regard the coordinate vectors as elements of a Jordan algebra
denoted as $J_2^{\mathbf C}$
\cite{mg75,mg80}.
  Then the rotation, Lorentz and conformal groups in
$d=4$ can be identified with the automorphism , reduced  structure
and M\"{o}bius ( linear fractional) groups of  the Jordan algebra of
 $2\times 2$
complex Hermitian matrices $J_2^{\mathbf C}$ \cite{mg80}.
 The reduced structure group $Str_0(J)$ of a Jordan algebra $J$ is
 simply the
invariance group of its norm form $N(J)$. (The structure group
 $Str(J)= Str_0(J)\times SO(1,1)$
,on the other hand,
is simply the invariance group of $N(J)$ up to an overall constant scale
factor.)
Furthermore, this
interpretation allows one to define generalized space-times whose
coordinates are parametrized by the elements of  Jordan algebras
\footnote{More generally one can define spacetimes coordinatized by the elements
of Jordan triple systems and study their symmetry groups \cite{mg92} . However,
in this talk we restrict ourselves to spacetimes (superspaces) coordinatized by
Jordan algebras ( Jordan superalgebras)} \cite{mg75}.
The  rotation $Rot(J)$,
Lorentz $Lor(J)$ and conformal $Con(J)$ groups of these generalized
 space-times are then identified with the automorphism $Aut(J)$,
 reduced structure $Str_0(J)$
and M\"{o}bius  M\"{o}(J) groups of the corresponding Jordan
 algebra \cite{mg75,mg80,mg91,mg92} \footnote{Similar algebraic
 structures appear also in the study of internal U-duality groups of
 extended supergravity theories \cite{sfmg,gkn}.}.
  Denoting as  $J_{n}^{\mathbf A}$ the Jordan algebra of $n\times
n$   Hermitian matrices over the division algebra ${\mathbf A}$ and the
Jordan algebra of Dirac gamma matrices in $d$ ( Euclidean) dimensions
as $\Gamma(d)$ one finds
the following symmetry groups of  generalized space-times defined by simple
Jordan algebras:

\begin{table}[ht]
 \caption{~}
\begin{center}
\begin{displaymath}
\begin{array}{|c|c|c|c|}
\hline
~&~&~&~\\
J & Rot(J) & Lor(J) & Con(J) \\
\hline
~&~&~&~\\
J_{n}^{\mathbf R} & SO(n) & SL(n,{\mathbf R}) & Sp(2n,{\mathbf R})\\
~&~&~&~\\
J_{n}^{\mathbf C} & SU(n) & SL(n,{\mathbf C}) & SU(n,n) \\
~&~&~&~\\
J_{n}^{\mathbf H} & USp(2n) & SU^{*}(2n) & SO^{*}(4n) \\
~&~&~&~\\
J_{3}^{\mathbf O} & F_{4} & E_{6(-26)} & E_{7(-25)} \\
~&~&~&~ \\
\Gamma(d) & SO(d) & SO(d,1) & SO(d,2) \\
~&~&~&~ \\
\hline
\end{array}
\end{displaymath}

\end{center}
\end{table}

The symbols ${\mathbf R}$, ${\mathbf C}$, ${\mathbf H}$, ${\mathbf O}$
 represent the four division
algebras.  For the Jordan algebras $J_n^{\mathbf A}$ the norm form is the determinantal
form ( or its generalization to the quaternionic and octonionic matrices).
For the Jordan algebra $\Gamma(d)$ generated by
 Dirac gamma matrices $\Gamma_{i} ~( i =1,2,...d)$

\begin{equation}
~\\~
\{\Gamma_i,\Gamma_j\} = \delta_{ij} {\mathbf 1};
~~~~~i,j,\ldots~=~ 1,2,\ldots,d \\
~\\
\end{equation}

the norm of a general element $x= x_0 {\mathbf 1} + x_i \Gamma_i$ of $\Gamma(d)$
is quadratic
and given by
\eq
N(x) = x \bar{x}= x_0^2 -x_ix_i
\en
where $\bar{x}= x_0 {\mathbf 1} - x_i \Gamma_i $.
Its automorphism, reduced structure and M\"{o}bius groups are
simply the rotation, Lorentz and conformal groups of
$(d+1)$-dimensional Minkowski spacetime. One finds the following
special isomorphisms between the Jordan algebras of $2\times 2$
Hermitian matrices over the four division algebras and the Jordan
algebras of gamma matrices:

\begin{equation}
~\\
J_{2}^{\mathbf R} \simeq \Gamma(2)~~~~;~~~~J_{2}^{\mathbf C} \simeq \Gamma(3) \\
~~~~;~~~~
J_{2}^{\mathbf H} \simeq \Gamma(5)~~~~;~~~~J_{2}^{\mathbf O} \simeq \Gamma(9) \\
~\\
\end{equation}

The Minkowski spacetimes they correspond to are precisely the
critical dimensions for the existence of super Yang-Mills
theories as well as of the classical Green-Schwarz superstrings.
These Jordan algebras are all quadratic and their norm forms are
precisely the quadratic invariants constructed using the
Minkowski metric.

\subsection{Covariant Quantum Fields over Generalized Spacetimes and the ULWR's of
Their Conformal Groups}

A remarkable fact about  Table 2 is that the maximal compact
subgroups of the generalized conformal groups of formally real Jordan algebras are
simply the compact forms of their structure groups
(generalized Lorentz group times dilatations). Furthermore, they all admit
unitary representations (positive energy) of the lowest weight type.
\footnote{ Similarly, the generalized conformal groups defined by
 Hermitian Jordan triple
systems all admit unitary irreducible representations of the lowest weight
type \cite{mg92}.}
 For example,  the conformal group of the Jordan algebra
$J_2^{\mathbf C}$ corresponding to the four dimensional Minkowski space
 is $SU(2,2)$  with a maximal compact subgroup
$SU(2)\times SU(2)\times U(1)$ which is simply the compact form of the
structure group $SL(2,\mathbf C)\times SO(1,1)$.
In \cite{gmz2} it was
explicitly shown how to go from the  compact
$SU(2)\times SU(2) \times U(1)$ basis of the ULWR's of $SU(2,2)$ to the
manifestly covariant  $SL(2,\mathbf C)\times SO(1,1)$ basis. The transition from
the compact  to the covariant basis corresponds simply to going
from a "particle" basis to a coherent state basis of the ULWR.
The coherent states are labelled by the elements of  $J_2^{\mathbf C}$
i.e by the coordinates of four dimensional Minkowski space. One can then
establish a one-to-one correspondence between irreducible ULWR's of
$SU(2,2)$ and the fields transforming irreducibly under the Lorentz
group $SL(2,{\mathbf C}) $ with a definite conformal dimension.
Thus one
can associate with irreducible ULWR's of $SU(2,2)$  fields transforming
covariantly under the Lorentz group with a definite conformal dimension.

Similarly, the conformal group $SO^*(8)$  of the Jordan algebra $J_2^{\mathbf H}$
parametrizing the six dimensional Minkowski space has a maximal compact subgroup
$U(4)$  which is the compact form of the
structure group $SU^*(4)\times SO(1,1)$. In \cite{mgst} it was shown explicitly
how to go from the  compact $U(4)$ basis of the ULWR's of $SO^*(8)$ to
the non-compact basis $SU^*(4) \times SO(1,1)$ which is simply the
Lorentz group in six dimensions times dilatations. The coherent states of the non-compact
basis are again labelled by the elements of $J_2^{\mathbf H}$, i.e the
coordinates of 6d Minkowski space.
Thus each irreducible
ULWR of $SO^*(8)$ can be identified with a field transforming
covariantly under the Lorentz group $SU^*(4)$ with a definite conformal
dimension.

The results obtained explicitly for the conformal groups of $J_2^{\mathbf C}$
and $J_2^{\mathbf H}$ extend to the conformal groups of all formally
real Jordan algebras and of Hermitian Jordan triple systems \cite{mg2000}.
The general theory can be summarized as follows: Let $g$ be the Lie
algebra of the conformal group of a formally real Jordan algebra and
$g^0$ the Lie algebra of its maximal compact subgroup. Then $g$ has
a three-graded decomposition with respect to $g^0$:
\begin{equation}
 g=g^- +g^0 + g^+  \nonumber
\end{equation}
where the grading is determined by the "conformal energy operator".
Now let $n^0$ be the Lie algebra of the structure group of the Jordan
algebra or triple system. Then $g$ has a 3-graded decomposition with
respect to $n^0$ as well:
\begin{equation}
  g=n^- + n^0 + n^+
\end{equation}
where the grading is defined by the generator of scale transformations.
In the compact basis an irreducible ULWR of $Conf(J)$ is uniquely determined by a lowest
weight vector $| \Omega \rangle $ transforming irreducibly under the
maximal compact subgroup $K$ that is annihilated by the operators belonging to $g^-$
\begin{equation}
  g^-| \Omega \rangle =0
\end{equation}

As was done explicitly for the conformal groups in 4 and 6 dimensions
\cite{gmz2,mgst} one can show that there exists a complex rotation
operator $W$ in the representation space with the property that
the vector $W| \Omega \rangle $ is annihilated by all the generators
belonging to $n^-$
\begin{equation}
  n^-W| \Omega \rangle =0
\end{equation}
and it transforms in a finite dimensional non-unitary representation of
the non-compact structure group. Remarkably the transformation properties of $W| \Omega \rangle $
under the structure group coincide with the transformation properties of
$| \Omega \rangle $ under the maximal compact subgroup $K$. In
particular, the conformal dimension of the vector $W| \Omega \rangle $
is simply the  negative of the conformal energy of $| \Omega \rangle $.
If one chooses a basis $e_{\mu}$ for the Jordan algebra $J$ and denote
the generators of generalized translations
in the space $n^+$ corresponding to $e_{\mu}$ as $P_{\mu}$, then the coherent
states defined by the action of generalized translations on $W| \Omega \rangle $
\begin{equation}
  |\Phi(x_{\mu} \rangle :=e^{i x^{\mu}P_{\mu}} W| \Omega \rangle
\end{equation}
form the covariant basis of the ULWR of the generalized conformal group
$Con(J)$ \footnote{We should note that the (super) coherent states associated
with ULWR's of non-compact (super) groups introduced in \cite{bg} are labelled
by complex (super) "coordinates" in the compact basis. These (super) coordinates
parametrize the (super) hermitian symmetric space $G/H$. }  .
The coherent states $|\Phi(x_{\mu} \rangle $ labelled by the coordinates
correspond to conformal fields transforming covariantly under the
Lorentz group with a definite conformal dimension. Since the state
$W| \Omega \rangle $ is annihilated by the generators of special
conformal transformations $K_{\mu}$ belonging to the space $n^-$ this
proves that the irreducible ULWR's are equivalent to  representations
induced by finite dimensional irreps of the Lorentz group with a
definite conformal dimension and trivial special conformal
transformation properties. This generalizes the well-known construction
of the positive  energy representations of the four dimensional
conformal group $SU(2,2)$ \cite{mack,gmz2} to all generalized conformal
groups of formally real Jordan algebras and Hermitian Jordan triple
systems. They are simply induced representations with respect to the
maximal parabolic subgroup $Str(J) \odot S_J$ where $ \odot$ denotes
semi-direct product and $S_J$ is the Abelian subgroup generated by
generalized special conformal transformations.

We should perhaps note that the generalized Poincar\'{e} groups associated
with the spacetimes defined by Jordan algebras are of the form
\begin{equation}
  \mathcal{PG}(J) := Lor(J) \odot T_J
\end{equation}
where $T_J$ is the Abelian subgroup generated by generalized
translations $P_{\mu}$.  For quadratic Jordan algebras, $\Gamma(d)$
, $\mathcal{PG}(\Gamma(d)) $ is simply the Poincar\'{e} group
in $d$ dimensional Minkowski space. The group $\mathcal{PG}(\Gamma(d)) $
has a quadratic Casimir operator $M^2 = P_{\mu} P^{\mu}$ which is simply
 the mass operator. For Jordan algebras $J$ of degree $n$ the
generalized Poincar\'{e} group $\mathcal{PG}(J)$ has a Casimir invariant
of order $n$ constructed out of the generalized translation generators
$P_{\mu}$. For example for the exceptional Jordan algebra $J_3^{\mathbf O}$
the corresponding Casimir invariant is cubic and has the form
\begin{equation}
M^3 = C_{\mu\nu\rho} P^{\mu}P^{\nu}P^{\rho}
\end{equation}
where $C_{\mu\nu\rho}$ is the symmetric invariant tensor of the
generalized Lorentz group $E_{6(-26)}$ of $J_3^{\mathbf O}$.

\section{Generalized superspaces defined by Jordan superalgebras
and their symmetry supergroups}
\setcounter{equation}{0}

The generalized space-times defined by Jordan algebras can be
extended to define generalized superspaces over Jordan superalgebras
and super Jordan triple systems \cite{mg80,mg91} .
 A Jordan superalgebra is a $Z_{2}$ graded algebra $J = J^{0}+J^{1}$ with a
supersymmetric product

\begin{equation}
\begin{array}{c}
~\\
a \cdot b~=~ (-1)^{\alpha\beta} b \cdot a\\
~\\
a~\in~J^{\alpha},~ ~b~\in J^{\beta}; ~ ~ \alpha,\beta = 0,1\\
~
\end{array}
\end{equation}

\noindent
which satisfies the identity

\begin{equation}
~\\
(-1)^{\alpha\gamma}[L_{a \cdot b},L_{c}\}~+~
(-1)^{\beta\alpha} [L_{b \cdot c},L_{a}\}~+~
(-1)^{\gamma\beta} [L_{c \cdot a},L_{b}\}~=~0 \\
~\\
\end{equation}
\noindent
where the mixed bracket [~,~\} denotes the usual Lie
superbracket and $L_a$ denotes left multiplication by the
element $a$ of $J$.
Jordan superalgebras  have been classified by Kac
\cite{vk}.

One defines the  generalized superspaces by
multiplying the even elements of a Jordan superalgebra $J$
by real coordinates and their odd elements by Grassmann coordinates
\cite{mg80,mg91}.
The  rotation, Lorentz and
conformal supergroups of these generalized superspaces are then
 given the  the automorphism, reduced structure
and M\"{o}bius supergroups of $J$. A complete list of these supergroups
was given in \cite{mg91}.   We reproduce  this list in Table 3.
\begin{table}[ht]
  \caption{Below we give generalized rotation, Lorentz and conformal supergroups
  of Jordan superalgebras $J$ by using a modified
version of Kac's notation for labeling Jordan superalgebras and giving
only the compact forms of the various supergroups. The Jordan superalgebra of
 type $X$ with $m$ even
elements and $n$ odd elements is denoted as $JX(m/n)$ and the
term $U(1)_{F}$ below denotes the ``fermionic" $U(1)$ factor generated
by a single odd generator.}
\begin{center}

\footnotesize
\begin{displaymath}
\begin{array}{|c|c|c|c|}
\hline
~&~&~&~\\
JX & SRot(JX) & SLor(JX) & SCon(JX) \\
\hline
~&~&~&~\\
JA(m^{2}+n^{2}/2mn) & SU(m/n) &
SU(m/n)^2 & SU(2m/2n) \\
~&~&~&~\\
JBC(\frac{m^2+m}{2}+2n^2-n/2mn) & OSp(m/2n) &
SU(m/2n) & OSp(4n/2m) \\
~&~&~&~\\
JD(m/2n) & OSp(m-1/2n) &
OSp(m/2n) & OSp(m+2/2n) \\
~&~&~&~\\
JP(n^{2}/n^{2}) & P(n-1) &
SU(n/n) & P(2n-1) \\
~&~&~&~\\
JQ(n^{2}/n^{2}) & Q(n-1)\times U(1)_{F} &
Q(n-1)^2 \times U(1)_{F} & Q(2n-1) \\
~&~&~&~\\
JD(2/2)_{\alpha} & OSp(1/2) &
SU(1/2) & D(2,1;\alpha) \\
~&~&~&~\\
JF(6/4) & OSp(1/2)\times OSp(1/2) &
OSp(2/4) & F(4) \\
~&~&~&~\\
JK(1/2) & OSp(1/2) &
SU(1/2) & SU(2/2) \\
~&~&~&~ \\
\hline
\end{array}
\end{displaymath}

\end{center}
\end{table}

\normalsize
The conformal groups of formally real Jordan algebras all admit ULWR's
and as explained above one can associate with each irreducible ULWR  a
 covariant conformal field with a definite conformal dimension.
 Hence we shall restrict
ourselves to those Jordan superalgebras or super Jordan triple systems
whose conformal supergroups admit unitary representations of the lowest
weight type. The general theory for the construction of  the unitary lowest weight
representations of non-compact supergroups was given in
\cite{bg} , both in a compact particle state basis as well as the compact
super-coherent state basis. The coherent states defined in
\cite{bg} for non-compact groups $G$ are
labelled by the complex variables parametrizing the hermitian symmetric
space $G/H$ where $H$ is the maximal compact subgroup. On the other hand
the coherent states defined in \cite{gmz2} for $SU(2,2)$ and in
\cite{mgst} for $OSp(8^*|4)$ as well as their generalizations to all
non-compact groups discussed in the previous section are labelled by
{\it real} (generalized) coordinates of the (generalized) space-times on which
$G$ acts as a (generalized) conformal group.

The even subgroup of (generalized) conformal supergroups $SCon(JX)$  are of the form
$G \times K$ where $G$ is the (generalized) conformal group and $K$ is
some compact internal symmetry group. The ULWR's of $SCon(JX)$ decompose
into a set of  irreducible ULWR's of $G\times K$. By acting on the
lowest weight vectors of the
irreducible ULWR's  of $G\times K$ with the operator
\begin{equation}
  e^{ix^{\mu} P_{\mu}} W
\end{equation}
one obtains a set of coherent states transforming covariantly under the
(generalized) Lorentz group $Lor(J)$ with definite conformal dimension.
Thus the irreducible ULWR's of $SCon(JX)$ correspond simply to a
supermultiplet of fields transforming irreducibly under $Lor(J) \times
K$ with definite conformal dimension. If one  starts from the
compact super-coherent state basis of $SCon(JX)$ and goes over to the
non-compact basis one obtains a "superstate" which corresponds to a
superfield built out of covariant fields multiplied by appropriate
Grassmann parameters \footnote{ Recently, a number of papers studied the
supermultiplets of conformal supergroups in 3,4 and 6 dimensions
using the formalism of superfields \cite{superfields}. Writing a ULWR
of a non-compact conformal supergroup as a superfield
corresponds to going to a super-coherent state basis for the
corresponding  ULWR in the oscillator formalism.} . It is also
possible to define covariant super-coherent states directly by acting on the
lowest weight vector $| \Omega \rangle$ of the ULWR of $SCon(JX)$ by
the operator
\begin{equation}
e^{ix^{\mu}P_{\mu}+ \theta^{\alpha} Q_{\alpha}} W
\end{equation}
where $Q_{\alpha}$ are the (generalized) "Poincar\'{e}" supersymmetry
generators. A detailed formulation of the covariant super-coherent state
basis of the ULWR's of generalized conformal supergroups will be given
elsewhere \cite{mg2000}.

Before concluding I should point out that the simple yet powerful oscillator method
for the construction of the
ULWR's of non-compact superconformal groups can be given a dynamical
realization in terms of twistorial or super-twistorial  fields
such that the (super)-oscillators become the Fourier modes of these
fields \cite{gmz2,cgkrz,ckr,mgst,mg2000}.


\newpage
\begin{thebibliography}{99}
\bibitem{mald} J. Maldacena, hep-th/9711200.
\bibitem{agmoo} O. Aharony, S. Gubser, J. Maldacena, H. Ooguri and
Y. Oz, hep-th/9905111.
\bibitem{mgdm} M. G\"{u}naydin and D. Minic, " Singletons, Doubletons and
M-Theory", hep-th/9702047.
\bibitem{gw} M.  G\"unaydin and N.P. Warner, Nucl. Phys. {\bf B 272} (1986) 99.
\bibitem{gnw} M. G\"unaydin, P. van Nieuwenhuizen and N.P. Warner, Nucl. Phys.
{\bf B 255} (1985) 63.
\bibitem{ptn} K. Pilch, P. K. Townsend and P. van Nieuwenhuizen,
Nucl. Phys. {\bf B242} (1984) 377.
\bibitem{gm}
M. G\"unaydin and N. Marcus, Class. Quantum Gravity {\bf 2} (1985)
L11.
\bibitem{krv} H. J. Kim, L. J. Romans and P. van Nieuwenhuizen.
Phys. Rev. {\bf D32} (1985) 389.
\bibitem{mgnm2} M. G\"unaydin and N. Marcus, Class. Quant. Grav. {\bf 2} (1985) L19.
\bibitem{pad} P.A.M. Dirac, J. Math. Phys. {\bf 4} (1963) 901.
\bibitem{cf1}
C. Fronsdal, Phys. Rev. {\bf D 12} (1975) 3819;\\
M. Flato and C. Fronsdal, Lett.
Math. Phys. {\bf 2} (1978) 421.
\bibitem{gs}
M. G\"unaydin and C. Saclioglu, Comm. Math. Phys., {\bf 87}
(1982)  159; Phys. Lett. {\bf B108} (1982) 180.
\bibitem{mg81} M. G\"{u}naydin, {\it Oscillator-like Unitary
Representations of Non-compact Groups and Supergroups and Extended
Supergravity Theories}, expanded version of the invited talk given
in {\it Group Theoretical Methods in Physics}, Istanbul, 1982, Ecole
Normale Superieure preprint LPTENS 83/5 and in Lecture Notes in Physics
Vol. 180  (1983), ed by E. In\"{o}n\"{u} and M. Serdaroglu.

\bibitem{gh}
M.  G\"unaydin and S.J. Hyun, J. Math. Phys. {\bf 29} (1988) 2367.
\bibitem{mg89}
M.  G\"unaydin, Nucl. Phys. {\bf B 6} (Proc. Suppl.) (1989) p. 140.
\bibitem{gmz1} M. G\"{u}naydin, D. Minic and M. Zagermann, hep-th/9806042,
Nucl. Phys. B534(1998) 96-120.
\bibitem{gmz2} M. G\"{u}naydin, D. Minic and M. Zagermann,
hep-th/9810226, Nucl. Phys. B544 (1999) 737-758.

\bibitem{mgst} M. G\"{u}naydin and S. Takemae, "Unitary Supermultiplets of
 $OSp(8^*|4)$ and the  $AdS_{7}/CFT_{6}$
Duality", hep-th/9910110

\bibitem{gst} M. G\"unaydin, G. Sierra and P.K. Townsend, Nucl. Phys. {\bf B 274}
(1986) 429.
\bibitem{mg90} M. G\"{u}naydin, in Proceedings of the Trieste
conference {\it "Supermembranes and Physics in 2+1 dimensions"},
eds. M. J. Duff, C. N. Pope and E. Sezgin, World Scientific,
1990, pp.442.
\bibitem{hc} Harish-Chandra, Am. J. Math. 77 (955) 743; ibid {\bf 78} (1956) 1.
\bibitem{bg}
I. Bars and M. G\"unaydin, Comm. Math. Phys., {\bf 91} (1983) 21.
\bibitem{mgrs}M. G\"unaydin and R. Scalise, Jour. Math. Phys. {\bf 32} (1991) 599.

\bibitem{mg88}
M. G\"unaydin, J. Math. Phys. {\bf 29} (1988) 1275.

\bibitem{mg75} M. G\"unaydin, Nuovo Cimento {\bf 29A} (1975) 467.
\bibitem{mg80}  M. G\"unaydin, Ann. Israel
Physical Society {\bf 3} (1980) 279.
\bibitem{mg91}  M. G\"unaydin, in "Elementary Particles and the Universe:
{\it Essays in Honor of Murray Gell-Mann}" , ed. by J.H. Schwarz ( Cambridge
University Press,  1991).

\bibitem{mg92} M. G\"{u}naydin,
``Generalized conformal and superconformal group actions and Jordan algebras,''
Mod.\ Phys.\ Lett.\  {\bf A8}, 1407 (1993)
[hep-th/9301050].


\bibitem{sfmg} S. Ferrara and M. G\"{u}naydin,
``Orbits of exceptional groups, duality and BPS states in string theory,''
Int.\ J.\ Mod.\ Phys.\  {\bf A13}, 2075 (1998)
[hep-th/9708025].


\bibitem{gkn} M. G\"{u}naydin, K. Koepsell and H. Nicolai, in
preparation.


\bibitem{mack} G. Mack and A. Salam, Ann.
Phys. 53 (1969) 174;  G. Mack, Comm. Math. Phys. 55 (1977) 1.

\bibitem{vk} V. Kac, Comm. in Algebra {\it 5} (1977), 1375.
\bibitem{superfields} L.~Andrianopoli, S.~Ferrara, E.~Sokatchev and B.~Zupnik,
hep-th/9912007 ;
~Ferrara and E.~Sokatchev,
hep-th/9912168 ;
S.~Ferrara and E.~Sokatchev,
hep-th/0001178 ;
hep-th/0003051.
\bibitem{mg2000} M. Gunaydin, in preparation.
\bibitem{cgkrz} P.Claus, M. Gunaydin, R. Kallosh, J. Rahmfeld and
Y. Zunger, JHEP 9905 (1999) 019, hep-th/9905112.
\bibitem{ckr} P. Claus, R. Kallosh
and J. Rahmfeld, hep-th/9906195.

\end{thebibliography}
\end{document}